\documentclass[twoside]{article}
\usepackage[utf8]{inputenc}
\usepackage[T1]{fontenc}
\usepackage{amsmath, amsfonts, amssymb}
\usepackage{mathtools}
\usepackage{times}
\usepackage{bm}
\usepackage{color}
\usepackage[natural, dvipsnames, pdftex]{xcolor}
\usepackage[pdftex, hidelinks]{hyperref}
\usepackage[round]{natbib}
\usepackage{multirow}
\usepackage[plain,noend]{algorithm2e}
\usepackage[top=2cm, bottom=2cm, left=3.5cm, right=3.5cm]{geometry}
\usepackage[calcwidth,  sf,  big, compact]{titlesec}
\titleformat{\section}
  {\normalfont\scshape \Large}{\thesection}{1em}{}
\usepackage{fancyhdr}
\usepackage{setspace}
\usepackage[amsmath,  thref,  thmmarks,  hyperref]{ntheorem}

\titleformat{\section}[block]{\centering \large \scshape }
{  {\thesection.}}{4pt}{   }
\titleformat{\subsection}[block]{\centering \large \itshape}
{  {\thesection.}}{4pt}{   }

\theoremheaderfont{\hskip\parindent\normalfont\scshape}
\theorembodyfont{\itshape}
\theoremseparator{.}

\theoremnumbering{arabic}
\newtheorem{theorem}{Theorem}
\theoremnumbering{arabic}

\theoremnumbering{arabic}

\theoremnumbering{arabic}

\theoremnumbering{arabic}

\theoremnumbering{arabic}
\theoremheaderfont{\hskip\parindent\normalfont\itshape}
\theorembodyfont{\rm}

\theoremnumbering{arabic}

\theoremnumbering{arabic}

\theoremnumbering{arabic}

\theoremnumbering{arabic}

\theoremnumbering{arabic}
\theoremheaderfont{\hskip\parindent\normalfont\itshape}
\theorembodyfont{\normalfont}

\font\QEDlogofont=msam10 at 10pt
\def\QEDlogo{\hbox{\QEDlogofont\char'003}}

\theoremsymbol{\QEDlogo}
\theoremheaderfont{\hskip\parindent\normalfont\itshape}
\theorembodyfont{\normalfont}
\theoremseparator{.}
\theoremstyle{nonumberplain}
\newtheorem{proof}{Proof}

\makeatletter
\renewcommand{\algocf@captiontext}[2]{#1\algocf@typo. \AlCapFnt{}#2} 
\def\@algocf@capt@plain{top}
\renewcommand{\algocf@makecaption}[2]{%
  \addtolength{\hsize}{\algomargin}%
  \sbox\@tempboxa{\algocf@captiontext{#1}{#2}}%
  \ifdim\wd\@tempboxa >\hsize
    \hskip .5\algomargin%
    \parbox[t]{\hsize}{\algocf@captiontext{#1}{#2}}
  \else%
    \global\@minipagefalse%
    \hbox to\hsize{\box\@tempboxa}
  \fi%
  \addtolength{\hsize}{-\algomargin}%
}
\makeatother

\newenvironment{tabnote}{\vskip7pt%
  \par\ignorespaces%
  }{\par}

\def\T{\top}
\def\IPT{{inverse probability of treatment }}

\newcommand{\E}{\mathrm{E}}
\renewcommand{\P}{\mathrm{pr}}
\newcommand{\V}{\mathrm{var}}
\renewcommand{\mid}{\,|\,}
\newcommand\independent{\protect\mathpalette{\protect\independenT}{\perp}}
\def\independenT#1#2{\mathrel{\rlap{$#1#2$}\mkern2mu{#1#2}}}
\newcommand{\notindep}{{\not\perp\!\!\!\!\perp}}

\def\d{{\,\textrm{d}}}
\def\and{\textsc{and }}
\newcommand{\email}[1]{\href{mailto:#1}{\texttt{#1}}}
\title{A Bayesian view of doubly robust causal inference}

\author{O. Saarela\thanks{Dalla Lana School of Public Health, University of Toronto, 155 College Street, Toronto, Ontario M5T 
3M7,
Canada. \email{olli.saarela@utoronto.ca}},\; L. R. Belzile\thanks{\'Ecole Polytechnique F\'ed\'erale de 
Lausanne, EPFL-SB-MATHAA-STAT,
Station 8, CH-1015
Lausanne, Switzerland. \email{leo.belzile@epfl.ch}}\; \and\;D. A. Stephens\thanks{Department 
of Mathematics and Statistics, McGill 
University, Montreal, Quebec H3A 2K6, Canada.
\email{d.stephens@math.mcgill.ca}}}
\date{}

\begin{document}
%




\maketitle

\begin{abstract}
In causal inference confounding may be controlled either through regression adjustment in an outcome model, or through propensity 
score adjustment or inverse probability of treatment weighting, or both. The latter approaches, which are based on modelling of 
the treatment assignment mechanism and their doubly robust extensions have been difficult to motivate using formal Bayesian 
arguments; in principle, for likelihood-based inferences, the treatment assignment model can play no part in inferences concerning 
the expected outcomes if the models are assumed to be correctly specified. On the other hand, forcing dependency between the 
outcome and treatment assignment models by allowing the former to be misspecified results in 
loss of the balancing property of the propensity scores and the loss of any double robustness. In this paper, we explain in the 
framework of misspecified models why doubly robust inferences cannot arise from purely likelihood-based arguments, and demonstrate 
this through simulations. As an alternative to Bayesian propensity score analysis, we propose a Bayesian posterior predictive 
approach for constructing doubly robust estimation procedures. Our approach appropriately decouples the outcome and treatment 
assignment models by incorporating the inverse treatment assignment probabilities in Bayesian causal inferences as importance 
sampling weights in Monte Carlo integration.

A revised version of this article has been accepted for 
publication in \textit{Biometrika}, published by Oxford University Press. \\
Saarela, O., Belzile, L. R. and D. A. Stephens. A Bayesian view of doubly robust causal inference, \textsl{Biometrika} (2016), 
103 (3): 667-681. \url{doi: 10.1093/biomet/asw025}.
\end{abstract}

\section{Introduction}

In causal inference contexts, confounding is most often controlled by one of two approaches: first, by conditioning on the 
confounders in a regression model for the outcome in an \textit{outcome regression} model; secondly, by modelling the treatment 
assignment mechanism to obtain so-called \textit{propensity score} values, and then using these scores to construct strata within 
the observed sample, or a pseudo-sample from a hypothetical population, within which treatment assignment is not confounded. This 
pseudo-sample can be obtained via inverse probability of treatment weighting of the original sample, analogously to survey 
sampling procedures. The outcome regression adjustment method requires correct specification of the regression function in order 
to obtain consistent inference; this may be achieved in practice using flexible regression strategies and complex functions of 
typically a large number of covariates.  The propensity score adjustment methods focus principally on the specification of the 
treatment assignment model, which may be similarly flexible or complex. Under either
approach,  sufficient control of confounding is therefore dependent on possibly unverifiable modelling assumptions.
This has motivated the development of \textit{doubly robust} methods in which both model components are specified,
but only one of them needs to be correctly specified to sufficiently control for confounding.

Adjustments that depend on the propensity score using regression or reweighting are not easy to interpret from the
Bayesian perspective, since Bayesian inferences are naturally based on modelling of the outcome, with modelling of
the treatment assignment playing no role in inference relating to the outcome/treatment relationship
\citep{robins:1997}. \citet{gustafson:2012} attempted a Bayesian interpretation as a compromise between a saturated outcome model
and a
parametric one; however, the treatment assignment model did not feature in this interpretation.
\citet{scharfstein:1999} and \citet{bang:2005} pointed out a connection between a doubly robust estimator and a
model-based estimator; the two are equivalent if the outcome model in the latter features the so called clever
covariate, a particular function of the propensity score.

Separately from the above developments, there has been a body of work studying Bayesian versions of propensity
score adjustment to control for confounding
\citep{hoshino:2008,mccandless:2009,mccandless:2009(2),mccandless:2010,mccandless:2012,an:2010,kaplan:2012,
zigler:2013,chen:2015,graham:2015}. These approaches require one of two kinds of compromises; they either force a
parametrization which makes the outcome and treatment assignment models dependent, thus losing the balancing property of the
propensity
score, or cut such a feedback in which case the inference procedures are no longer Bayesian. Because of
these difficulties, some authors \citep[e.g.][p. 828]{achybrou:2010} have been content to fix the propensity
scores to their best estimates in model-based inferences, without attempting to jointly estimate the two model components.  In an 
alternative approach, \citet{wang:2012} and \citet{zigler:2014} have suggested connecting the outcome and treatment assignment 
models through the prior distribution in order to incorporate the uncertainty in confounder selection.
Herein we do not consider model uncertainty, but rather,
concentrate on inferences with a priori specified outcome and treatment assignment models.

The purpose of this paper is to clarify the theoretical and practical motivations for Bayesian propensity score
adjustment, and the relationships between the different methods proposed for this, which have not been fully
explored previously. We address these in the context of double adjustment for both the potential confounders and the propensity 
score, and argue that the problem cannot be properly understood
without considering it in the framework of misspecified models. To provide an alternative to Bayesian propensity
score adjustment, we propose deriving Bayesian versions of various \IPT weighted estimators, including \IPT weighted
outcome regression and the semi-parametric double robust estimator, through posterior predictive expectations,
with the weights introduced as importance sampling weights in Monte Carlo integration.

\section{Preliminaries}\label{section:preliminaries}

\subsection{Notation and assumptions}\label{section:notation}

For simplicity, we consider the case of a single binary treatment, and defer discussion of the longitudinal,
multiple exposure and continuous cases to the Discussion. Let the random vectors
$X_i$ represent a set of pre-treatment covariate measurements, $Z_i$ a binary
treatment allocation indicator, and $Y_i$ an outcome for individual $i$, measured
after sufficient time has passed since administering the treatment.
We adopt, for convenience, the standard \textit{potential outcome}
(or \textit{counterfactual}) framework: for individual $i$, the observed outcome is related to the two possible
potential outcomes $(\mathbf Y_{0i}, \mathbf Y_{1i})$ by $Y_i = (1 - Z_i) \mathbf Y_{0i} + Z_i \mathbf Y_{1i}$.
We assume that $X_i$ includes a sufficient set of covariates to control for confounding
in the sense that $Z_i \independent (\mathbf Y_{0i}, \mathbf Y_{1i}) \mid X_i$ (cf. ignorable treatment
assignment,
\citealp[][p. 43]{rosenbaum:1983}). The \textit{propensity score} $e(X_i) \equiv \P(Z_i = 1 \mid X_i)$
has the balancing property $Z_i \independent X_i \mid e(X_i)$,
which also implies that $(\mathbf Y_{0i}, \mathbf Y_{1i}) \independent Z_i \mid e(X_i)$ --
this scalar score is therefore useful in controlling for confounding.

If the covariate space $X_i$ is high-dimensional, in practice the task of controlling for confounding
often involves some covariate selection; here one can either select for the features that
predict the outcome, or the treatment assignment. To represent this,
let $S_i$ and $B_i$ represent some a priori selected subsets of the all observed features $X_i$,
so that the latter can be partitioned as $X_i = (S_i, R_i)$ or $X_i = (B_i, C_i)$,
where possibly $R_i = \emptyset$ and $C_i = \emptyset$.
If the selected set of features $S_i$ captures all relevant prognostic information,
then $\mathbf Y_{0i} \independent X_i \mid S_i$ \citep{hansen:2008}. For the remainder
of the paper, we consider the stronger condition (i) $(\mathbf Y_{0i}, \mathbf Y_{1i}) \independent X_i \mid S_i$,
which requires that $S_i$ also captures all relevant information about possible effect modification.

In this case $S_i$ is sufficient to control for confounding, since from the
properties of conditional independence \citep{dawid:1979} it follows that $(\mathbf Y_{0i}, \mathbf Y_{1i})
\independent Z_i \mid
S_i$.
If, on the other hand, the selected set of features $B_i$ has the balancing property
(ii) $Z_i \independent X_i \mid B_i$, it is sufficient to control for confounding.
We are interested in estimation procedures which are valid when either
$(\mathbf Y_{0i}, \mathbf Y_{1i}) \independent X_i \mid S_i$
(but possibly $Z_i \notindep X_i \mid B_i$) or
$Z_i \independent X_i \mid B_i$ (but possibly $(\mathbf Y_{0i}, \mathbf Y_{1i}) \notindep X_i \mid
S_i$).

Our parameter of interest is an average causal contrast such as
\[\E(\mathbf Y_{1i}) - \E(\mathbf Y_{0i}) = \E_{X_i}\{\E(\mathbf Y_{1i} \mid X_i)\} - \E_{X_i}\{\E(\mathbf Y_{0i}
\mid X_i)\}.\]
Interest then lies in the identifiability of the average potential outcomes based on the observed
data. When the  `no unmeasured confounder' assumption holds, it follows that \citep[e.g.][p. 43]{hernan:2006}
\begin{equation*}
\E(\mathbf Y_{1i}) - \E(\mathbf Y_{0i}) =
\int_{x_i} \left\{ \E(Y_i \mid Z_i = 1, x_i) - \E(Y_i \mid Z_i = 0, x_i) \right\} p(x_i) \d x_i.
\end{equation*}

\subsection{Bayesian model formulation for outcome regression}\label{section:Bayesframework}

Any Bayesian model specification is constructed via de Finetti's representation for exchangeable random sequences
$v_i \equiv (x_i, y_i, z_i)$ (see for example \citealp[][Chap. 4]{bernardo:1994}). The (subjective) joint
distribution
for observations $v \equiv (v_1, \ldots, v_n)$ may then be represented by
\begin{align}\label{eq:representation}
p(v) & = \int_{\phi, \gamma, \psi} \left\{\prod_{i=1}^n
p(y_i \mid z_i, x_i; \phi) p(z_i \mid x_i; \gamma) p(x_i; \psi) \right\} \pi_0(\phi, \gamma, \psi) \d\phi \d\gamma
\d\psi,
\end{align}
implying the existence of the parametric models and the prior density $\pi_0(\phi, \gamma, \psi)$.
Since the representation theorem is not constructive, that is, does not specify the models implicit in \eqref{eq:representation},
in practice inferences about a given finite-dimensional parametrization
involves the often implicit assumption that
$p(y_i \mid z_i, x_i; \phi_0) = f(y_i \mid z_i, x_i)$, $p(z_i \mid x_i; \gamma_0) = f(z_i \mid x_i)$
and $p(x_i; \psi_0) = f(x_i)$, where $(\phi_0, \gamma_0, \psi_0)$ is the limiting value
of the posterior $\pi_n(\phi, \gamma, \psi) \equiv p(\phi, \gamma, \psi \mid v)$ in the sense of
e.g. \citet[][p. 139]{vandervaart:1998}, and where the $f$s represent the `true' limiting (sampling) distributions.
We might further assume that the parameters are a priori independent,
so that $\pi_0(\phi, \gamma, \psi) = \pi_0(\phi)\pi_0(\gamma)\pi_0(\psi)$, in which case it
it also follows also that the posterior factorizes as
$\pi_n(\phi, \gamma, \psi) = \pi_n(\phi)\pi_n(\gamma)\pi_n(\psi)$ \citep[e.g.][p. 354--355]{gelman:2004}.

The marginal distribution $p(x_i; \psi)$ can in practice be specified nonparametrically
and estimated using the empirical covariate distribution, leading to a Bayesian
estimator of the average causal contrast, 
\begin{align}\label{equation:ppredictive}
\int_\phi \frac{1}{n} \sum_{i=1}^n \left\{ m(1, x_i; \phi) -
m(0, x_i; \phi) \right\} \pi_n(\phi) \d \phi,
\end{align}
where $\pi_n(\phi) \propto \textstyle \prod_{i=1}^n p(y_i \mid z_i, x_i; \phi)\pi_0(\phi)\d \phi$, and
$m(z,x;\phi) \equiv \int y p(y \mid z, x; \phi)\d y$. In Supplementary Appendix 1, we show that
\eqref{equation:ppredictive} can be motivated without the use of potential outcomes notation
through posterior predictive expectations for a new observation under a hypothetical completely randomized setting.

The estimator \eqref{equation:ppredictive} is the Bayesian version of the well-known direct standardization
or \emph{g}-formula. We return to it in Section \ref{section:iptw}, but note here that it does not feature the treatment 
assignment model;
rather, the posterior predictive approach for estimating the marginal causal contrast depends entirely on correct
specification of the distribution $p(y_i \mid z_i, x_i; \phi)$, or in a moment-based representation,
$m(z,x;\phi)$. Before discussing Bayesian alternatives to \eqref{equation:ppredictive}, we briefly review
some commonly used frequentist approaches for combining outcome regression and propensity score adjustment.

\section{Frequentist approaches for combining outcome regression and propensity score
adjustment}\label{section:frequentist}

\subsection{Including propensity scores into outcome regression}\label{section:psadjustment}

Because of the balancing property of the propensity score, it is tempting to specify a
propensity score $e(b_i; \gamma) \equiv \P(Z_i = 1 \mid b_i; \gamma)$ and use a statistical model such as
$p\{y_i \mid z_i, e(b_i), s_i; \phi\}$, in the hope that, if the prognostic model is
is misspecified, adjusting for the propensity score would still sufficiently control for any residual confounding.
For simplicity, we take the parameters $\phi$ to specify also the functional dependence between the propensity
score and the outcome; to model this dependency, it is advisable to use flexible formulations
such as splines \citep[e.g.][]{zhang:2009}. Using such an outcome model, the marginal causal contrast
would then be estimated by
\begin{align}\label{equation:psadjustment}
\frac{1}{n} \sum_{i=1}^n \left[ m \left\{1, e(b_i; \widehat\gamma),s_i ; \widehat\phi\right\}
- m\left\{0, e(b_i; \widehat\gamma),s_i ; \widehat\phi\right\} \right],
\end{align}
where $m\{z, e(b; \gamma),s ; \phi\} \equiv \int y p\{y \mid z, e(b; \gamma), s; \phi\}\d y$, and
where $\widehat\phi$ and $\widehat\gamma$ are maximum likelihood estimators for the parameters
in the outcome regression and treatment assignment model, respectively.
The motivation for such a double adjustment is that it is sufficient to control for confounding
if either condition (i) or (ii) of Section \ref{section:notation} applies. We summarize this property
in the following theorem (a proof in Supplementary Appendix 2; in the notation all convergencies are in probability unless
otherwise stated).

\begin{theorem}\label{doublerobustness}
Estimator \eqref{equation:psadjustment} is consistent if the outcome model is correctly specified
in the sense that $m\{z, e(b; \gamma_0),s ; \phi_0\} = \int y f(y \mid z, e(b), s)\d y$,
the parameters in this can be consistently estimated so that $\widehat\phi \rightarrow \phi_0$,
the treatment assignment model is correctly specified in the sense that $p(z_i \mid b_i; \gamma_0) = f(z_i \mid
b_i)$ and
$\widehat\gamma \rightarrow \gamma_0$, and either (i) holds true, or (ii) holds true.
\end{theorem}

The estimator \eqref{equation:psadjustment} may be considered `doubly robust'
in terms of the covariate selection in the sense that only one of the sets $S_i$ and $B_i$ needs to be correctly
specified, although it still always relies on a correct parametric specification of the model for the expected
outcome, conditional on $\{Z_i, e(B_i), S_i\}$.

\subsection{The clever covariate and augmented outcome regression}\label{section:cc}

The estimator discussed in the previous section did not specify
which function of the propensity score $e(B_i)$ should be added to the regression model.
\citet[][p. 1141--1142]{scharfstein:1999} and \citet[][p. 964--965]{bang:2005}
drew a connection between propensity score regression adjustment and doubly robust estimators of the form
\begin{align*}
\MoveEqLeft\frac{1}{n} \sum_{i=1}^n 
\frac{y_i z_i - \{z_i - e(b_i; \widehat\gamma)\} m(1,s_i;\widehat \phi)}{e(b_i;
\widehat\gamma)} \\&\quad -\frac{1}{n} \sum_{i=1}^n
\frac{y_i (1 - z_i) - [(1 - z_i) - \{1 - e(b_i; \widehat\gamma)\}] m(0,s_i;\widehat \phi)}{1 - e(b_i;
\widehat\gamma)}, 
\end{align*}
which can be equivalently represented as
\begin{align}
\frac{1}{n} \sum_{i=1}^n \bigg\{ \frac{z_i}{e(b_i; \widehat\gamma)} - \frac{1 - z_i}{1 - e(b_i; \widehat\gamma)}
\bigg\}
\left\{y_i - m(z_i,s_i;\widehat \phi)\right\} + \frac{1}{n} \sum_{i=1}^n \bigg\{m(1,s_i;\widehat \phi) -
m(0,s_i;\widehat \phi) \bigg\} \label{eq:BangRobins}.
\end{align}
On considering the score equation derived from a regression of $Y_i$ on $Z_i$ and $S_i$ with mean function
$m(z,s;\phi)$, this
form suggests incorporating the derived covariate
$c(z_i, b_i) = z_i/e(b_i) - (1 - z_i)/\{1 - e(b_i)\}$
\citep[termed the `clever covariate' by][p. 8]{rose:2008} additively into the outcome regression, that is, for
example
\begin{equation}\label{eq:clevercovariate}
m(z,s;\phi) = \phi_0 + \phi_1 z + \phi_2^\T s + \phi_3 c(z, b).
\end{equation}
The first term in \eqref{eq:BangRobins} is then zero through the maximum likelihood score equation, leaving only
the last term which is the model based estimator of the marginal treatment effect. Thus with the clever covariate
in the outcome model, the doubly robust estimator is equivalent to the model-based estimator.
In the special case of model \eqref{eq:clevercovariate}, this becomes
\begin{align*}
\frac{1}{n} \sum_{i=1}^n \left\{m(1,x_i;\widehat \phi) - m(0,x_i;\widehat \phi) \right\}
=
\widehat\phi_1 + \widehat\phi_3 \frac{1}{n} \sum_{i=1}^n \left\{\frac{1}{e(b_i; \widehat\gamma)} - \frac{1}{1 -
e(b_i; \widehat\gamma)}\right\}.
\end{align*}
A potential drawback of using this covariate is that it
may lead to extreme variability for the resulting mean difference estimator,
even compared to inverse probability of treatment weighted estimators of the form
\begin{align} \label{eq:iptw}
\frac{1}{n} \sum_{i=1}^n \bigg\{ &\frac{y_i z_i}{e(b_i; \widehat\gamma)} - \frac{y_i (1 - z_i)}{1 - e(b_i;
\widehat\gamma)}\bigg\}.
\end{align}
To see why this is the case, 
the distribution of $c(z_i, b_i)$ in itself can be very skewed to the
right, but this becomes even more pronounced in the model-based estimator where the clever covariate has to
evaluated
at both $c(1, b_i)$ and $c(0, b_i)$ for each $i = 1, \ldots, n$. In contrast,
the \IPT weighted estimator \eqref{eq:iptw} involves only the probabilities of treatments that were actually
assigned.

Eq.~\eqref{eq:BangRobins} is doubly robust in a stronger, semi-parametric, sense than
\eqref{equation:psadjustment}; it does not require correct specification of the outcome model,
if the treatment assignment model is correctly specified.
The approximate Bayesian double robust approach proposed by \citet{graham:2015}
involved replacing $m(z,x;\phi)$ in \eqref{equation:ppredictive} with a
linear predictor augmented with the clever covariate. We take this to be a special
case of the two-step Bayesian methods to be discussed in Section \ref{section:unknown},
and thus do not separately consider it in the present paper. However, in Section
\ref{section:isdr} we will show how the form \eqref{eq:BangRobins} may be derived
through posterior predictive expectations and importance sampling.

\subsection{Inverse probability of treatment weighted outcome regression}\label{section:iptwor}

Yet another estimator for the marginal causal contrast is
\begin{align}\label{equation:standardization}
\frac{1}{n} \sum_{i=1}^n \left\{ \E(\mathbf Y_{1i} \mid s_i; \widehat\phi)
- \E(\mathbf Y_{0i} \mid s_i; \widehat\phi)\right\},
\end{align}
where the parameters $\phi$ in the model for the potential outcomes $(\mathbf Y_{1i}, \mathbf Y_{0i})$
are estimated using an \IPT weighted estimating function $l(\phi) = \sum_{i=1}^n l_i(\phi)$, where
\begin{align*}
l_i(\phi) = \sum_{a=0}^1 \mathbf 1_{\{z_i = a\}} \frac{\log p(\mathbf y_{ai} \mid s_i; \phi)}{\P(Z_i = a \mid
b_i)}.
\end{align*}
The corresponding estimating equation is $u(\phi) = \sum_{i=1}^n u_i(\phi) = 0$, where the pseudo-score function is
\begin{align}\label{equation:score}
u_i(\phi) &= \sum_{a=0}^1 \mathbf 1_{\{z_i = a\}}\frac{\textstyle \frac{\partial}{\partial \phi}\log p(\mathbf
y_{ai} \mid s_i; \phi)}{\P(Z_i = a \mid b_i)}.
\end{align}
Here the treatment assignment probabilities $\P(Z_i = a \mid b_i)$
would in practice be replaced with estimates $\P(Z_i = a \mid b_i; \widehat\gamma)$.
To motivate the use of \eqref{equation:score} for the parameter estimation, we present the following theorem (a proof given in
Supplementary Appendix 2).

\begin{theorem}\label{eequation}
The estimating equation $u(\phi) = 0$ is unbiased
if the model for the potential outcomes is correctly specified in the sense
that $p(\mathbf y_{ai} \mid s_i; \phi_0) = f(\mathbf y_{ai} \mid s_i)$,
and either (i) or (ii) holds true.
\end{theorem}

If we further assume the consistency of the estimator for $\phi$,
as well as consistency of $\widehat\gamma$ when the weights are correctly specified,
it also follows that \eqref{equation:standardization} consistently estimates the marginal causal contrast.
In Section \ref{section:isor} we demonstrate that an estimator of the form
\eqref{equation:standardization} can also be motivated from Bayesian arguments.

\section{Two-step estimation when the propensity score is unknown}\label{section:unknown}

In observational settings the function $e(B_i)$ is unknown and has to be estimated.
When using an estimator of the form \eqref{equation:psadjustment}, a
central question from a Bayesian perspective then is how the uncertainty in the estimation of the parameters
$\gamma$ is incorporated in the inference of the marginal causal contrast. A `Bayesian' approach could be
motivated
by writing the posterior predictive expectation as
\begin{align}\label{eq:ppredictive1}
\int_{\psi, \gamma, \phi} \int_{x_{i}}
m(a,e(b_{i};\gamma),s_{i};\phi) p(x_{i}; \psi) \pi_n(\phi \mid \gamma) \pi_n(\gamma) \pi_n(\psi)
\d x_{i} \d \phi \d \gamma \d \psi,
\end{align}
where
\begin{align*}
\pi_n(\phi \mid \gamma) \propto \prod_{i=1}^n p\{y_{i} \mid z_{i}, e(b_{i}; \gamma), s_{i}; \phi\} \pi_0(\phi)
\quad\textrm{ and }\quad \pi_n(\gamma) \propto \prod_{i=1}^n p(z_i \mid b_i; \gamma) \pi_0(\gamma).
\end{align*}

The integrals of the form \eqref{eq:ppredictive1} could be evaluated by Monte Carlo integration, by forward
sampling first from $\pi_n(\gamma)$ and given the current value $\gamma$, from the conditional posterior
$\pi_n(\phi \mid \gamma)$. However, a concern related to such an approach is that
the product of the posterior distributions $\pi_n(\phi \mid \gamma)$ and
$\pi_n(\gamma)$ in \eqref{eq:ppredictive1} does not necessarily
correspond to any well defined joint posterior, except in the case
when the outcome is correctly specified in the sense (i), in which case it does not depend on the propensity score.
In contrast, for the correctly specified models in \eqref{eq:representation} we indeed have that
$p(\phi \mid v)p(\phi \mid x, z) = p(\phi, \gamma \mid v)$.
As a result, the above outlined two-step approach is not proper Bayesian,
and would have to be evaluated on its frequency-based properties.

To give an example of a situation where the two-step Bayesian approach does not result
in correct frequency-based inferences, we can consider the special case of the outcome model
$m\{z,e(b;\gamma),s;\phi\} =
\phi_0 + \phi_1 z + \phi_2^\T s + \phi_3^\T g\{e(b; \gamma)\}$,
where $g$ is, for example, some appropriate spline basis transformation of the propensity score.
With this model the estimator based on \eqref{eq:ppredictive1} for the average causal contrast
$\E(\mathbf Y_{1i}) - \E(\mathbf Y_{0i})$ reduces to
$\int_{\phi, \gamma} \phi_1 \pi_n(\phi \mid \gamma) \pi_n(\gamma) \d\phi \d\gamma$,
that is, to an estimator of the posterior mean $\E_{\Gamma \mid X, Z}\left\{\E(\Phi_1 \mid v; \gamma)\right\}$.
This estimator in turn can be approximated with $\sum_{j=1}^m \widehat \phi_1(\gamma^{(j)})/m$,
where $(\gamma^{(1)}, \ldots, \gamma^{(m)})$ is a Monte Carlo sample from
$\pi_n(\gamma)$ \citep[cf.][p. 589]{kaplan:2012}. We note first
that this estimator has the same asymptotic distribution as
$\widehat \phi_1(\widehat \gamma)$, where the treatment assignment model parameters have been fixed
to their maximum likelihood estimates (see Supplementary Appendix 3). In Supplementary Appendix 3 we further show that
$\mathrm{avar}\{\widehat\phi_1(\widehat \gamma)\} \le \mathrm{avar}\{\widehat\phi_1(\gamma_0)\}$, where
$\widehat\phi_1(\gamma_0)$ is the estimator given the `true' propensity scores.
Thus, with the propensity score adjusted outcome model specification,
a variance adjustment due to estimating the propensity scores should
reduce the asymptotic variance of the resulting treatment effect estimator compared to
a hypothetical situation where the true propensity scores are known (\citealp[cf.][]{henmi:2004}).
In contrast, \citet[p. 592]{kaplan:2012} and \citet[p. 11]{graham:2015}
propose variance estimators based on the variance decomposition formula
\begin{align}\label{equation:miformula}
\V(\Phi_1 \mid v) &= \E_{\Gamma \mid X, Z}\left\{\V(\Phi_1 \mid v; \gamma)\right\} + \V_{\Gamma \mid X,
Z}\left\{\E(\Phi_1 \mid v; \gamma) \right\},
\end{align}
which appears to add a further variance component. An explanation for the discrepancy is that
with the correctly specified models in the representation \eqref{eq:representation},
we have that $p(\phi_1 \mid v; \gamma) = p(\phi_1 \mid v)$, and the the second variance component becomes zero.
On the other hand, if the models are misspecified, the product form likelihood in the representation
\eqref{eq:representation} does not apply in the first place.
This illustrates the difficulty in applying Bayesian procedures in the context of
incompatible models. Even though this is routinely done in the context of multiple
imputation \citep[e.g.][]{rubin:1996}, and often produces reasonable results, in the present context there is
little motivation to use an approach which introduces an additional variance component to the
posterior variance, given that estimation of the propensity scores should reduce the variance
of the treatment effect estimator. We further discuss this discrepancy in the following section.

\section{Joint estimation of outcome and treatment assignment models}\label{section:jointestimation}

Even though the outcome $Y_i$ can obviously be predictive of the individual treatment assignment $Z_i$,
the outcomes are not informative of the treatment assignment mechanism (a proof in Supplementary Appendix 2):
\begin{theorem}\label{noninformative}
If the outcome model is correctly specified, the outcomes are non-informative of the parameters characterizing
the treatment assignment process.
\end{theorem}

In such a case the treatment assignment model plays no part in the inferences, since
the corresponding posterior predictive estimator is \eqref{equation:ppredictive}.
However, the Bayesian propensity score approach proposed by \citet{mccandless:2009} specifies
a parametrization making the outcome and treatment assignment models dependent and estimates
the parameters jointly. More recently, \citet{zigler:2013} suggested that a similar approach
could be used to obtain a Bayesian analogue to doubly robust inferences. 
Such an approach can be understood by assuming that there exists
a de Finetti parametrization $(\phi^*, \gamma^*)$ for which
\begin{equation*}
p(v) = \int_{\phi^*, \gamma^*, \psi} \left\{\prod_{i=1}^n
p(y_i, z_i \mid x_i; \phi^*, \gamma^*) p(x_i; \psi) \right\} \pi_0(\phi^*)\pi_0(\gamma^*)\pi_0(\psi) \d\phi^*
\d\gamma^*
\d\psi,
\end{equation*}
where $p(y_i, z_i \mid x_i; \phi^*, \gamma^*) = p\{y_i \mid z_i, e(b_{i}; \gamma^*), s_{i}; \phi^*\}
p(z_i \mid b_i; \gamma^*)$.
Compared to $(\phi, \gamma)$ in \eqref{eq:representation}, neither $\phi^*$ or $\gamma^*$ retains the original
interpretation, but now there is a well defined joint posterior distribution
\begin{align*}
\pi_n(\phi^*, \gamma^*) \propto
\prod_{i = 1}^n \left[p\{y_{i} \mid z_{i}, e(b_{i}; \gamma^*), s_{i}; \phi^*\} p(z_i \mid b_i; \gamma^*) \right]
\pi_0(\phi^*)\pi_0(\gamma^*).
\end{align*}
Inferences could now be based on the posterior predictive expectations
\begin{align}\label{eq:ppredictive3}
\int_{\phi^*, \gamma^*, \psi} \int_{x_{i}} m\{a,e(b_{i};\gamma^*),s_{i};\phi^*\} p(x_{i}; \psi)
\pi_n(\phi^*, \gamma^*) \pi_n(\psi) \d x_i \d\phi^* \d\gamma^* \d\psi .
\end{align}
At first sight, \eqref{eq:ppredictive3} would seem more natural than \eqref{eq:ppredictive1},
since the specification \eqref{eq:ppredictive3} does not make use of incompatible models.
However, now the quantities $e(b_{i}; \gamma^*)$ do not possess the balancing properties of propensity scores,
and thus it would be difficult to show whether \eqref{eq:ppredictive3} would have
the `double robustness' property of the estimator \eqref{equation:psadjustment}.

To address the lack of balance, \citet{mccandless:2010}
suggested a Gibbs sampler type approach similar to that of \citet{lunn:2009}
to cut the feedback from the outcome model,
by successively drawing from the conditional posteriors
$\pi_n(\gamma)$ and $\pi_n(\phi \mid \gamma)$ to approximate the joint posterior
of $(\phi^*, \gamma^*)$. However, as discussed in the previous Section, these posteriors are incompatible and
generally such a sampling procedure is not guaranteed to converge to any well defined joint distribution.
In fact, if the conditional posteriors can be sampled directly, or if the second sampling step
is allowed to converge to the corresponding conditional distribution, the inferences based on
the formulations \eqref{eq:ppredictive1} and \eqref{eq:ppredictive3} will be equivalent.

To sum up, trying to recover fully probabilistic inferences through sampling from a joint posterior
of the outcome and treatment assignment model parameters loses the balancing property of the propensity scores,
and consequently, the properties of the resulting estimator would be difficult to establish.
On the other hand, cutting the feedback in an attempt
to recover the balancing property would mean that the inferences are no longer based on well-defined
posterior distributions. Thus, in the following section,
following the approach outlined in \citet{saarela:2015a},
we formulate alternative Bayesian estimators that are not based on Bayesian propensity score adjustment.

\section{Posterior predictive inferences with importance sampling}\label{section:iptw}

\subsection{Inverse probability of treatment weighted outcome regression}\label{section:isor}

It has been recognized by various authors \citep[e.g.][]{roysland:2011,chakraborty:2013} that
\IPT weighting can be motivated through a change of probability measures, or equivalently,
importance sampling. However, as far as we know, before \citet{saarela:2015a}
this approach has not been used to formulate Bayesian causal inferences. Here we argue that
this approach can be used to resolve the paradoxes discussed in Sections \eqref{section:unknown}
and \eqref{section:jointestimation}. We follow the posterior predictive reasoning of Supplementary Appendix 1,
but rather than trying to directly predict a new observation
under the experimental setting $\mathcal E$, we consider first the task of parameter estimation
under this regime. For this purpose, a Bayes estimator for the outcome model
parameters can be constructed by maximizing the expected utility
$\E_{\mathcal E}\{l(\phi; V_i) \mid v\}$ with respect to  $\phi$, where the log-likelihood
$l(\phi; v_i) \equiv \log p(y_i \mid z_i, s_i; \phi)$ takes the role of a parametric
utility function, and the expectation is over a predicted new observation $v_i = (y_i, z_i, x_i)$, $i \notin \{1,
\ldots, n\}$.

Let further $\xi$ be a set of parameters characterizing the entire data-generating
mechanism under the observational regime $\mathcal O$. We can further write the
expectation as \[\E_{\mathcal E}\{l(\phi; V_i) \mid v\} = \E_{\Xi \mid V}\left[\E_{\mathcal E}\{l(\phi; V_i) \mid
v; \xi\}\right],\]
where, following \citet[][p. 26--27]{walker:2010}, we can consider the lower dimensional decision of maximizing
the expected utility $\E_{\mathcal E}\{l(\phi; V_i) \mid v; \xi\}$ with respect to  $\phi$ conditional on $\xi$.
With a known regime $\mathcal E$ and the stability assumption discussed in Supplementary Appendix 1,
$\arg\max_{\phi}\E_{\mathcal E}\{l(\phi; V_i) \mid v; \xi\}$ is a deterministic function of $\xi$.
Thus, the uncertainty represented by the posterior distribution $p_{\mathcal O}(\xi \mid v)$ then also reflects
the uncertainty
on $\phi$, providing means to construct a posterior distribution for $\phi$. This proceeds as follows; the
inner expectation can be written as
\begin{align*}
\E_{\mathcal E}[l(\phi; V_i) \mid v; \xi]
&= \int_{v_i} l(\phi; v_i) p_{\mathcal E}(v_i \mid v; \xi) \,\textrm dv_i \\
&= \int_{v_i} l(\phi; v_i) \frac{p_{\mathcal E}(v_i \mid v; \xi)}{p_{\mathcal O}(v_i \mid v; \xi)} p_{\mathcal
O}(v_i \mid v;
\xi)
\,\textrm dv_i \\
&= \int_{v_i} l(\phi; v_i) \frac{p_{\mathcal E}(y_i \mid z_i, x_i, v; \xi)p_{\mathcal E}(z_i)p_{\mathcal E}(x_i
\mid v; \xi)}
{p_{\mathcal O}(y_i \mid z_i, x_i, v; \xi)p_{\mathcal O}(z_i \mid x_i, v; \xi)p_{\mathcal O}(x_i \mid v; \xi)}
p_{\mathcal O}(v_i \mid v;
\xi)
\,\textrm dv_i \\
&= \int_{v_i} l(\phi; v_i) \frac{p_{\mathcal E}(z_i)}
{p_{\mathcal O}(z_i \mid x_i, v; \xi)} p_{\mathcal O}(v_i \mid v; \xi) \,\textrm dv_i,
\end{align*}
where in the last equality we made use of the stability assumption of Supplementary Appendix 1.
In the last form we can replace the predictive distribution under $\mathcal O$ with the
Bayesian bootstrap specification $p_{\mathcal O}(v_i \mid v; \xi) = \textstyle \sum_{k=1}^n \xi_k
\delta_{v_k}(v_i)$,
where $\xi \equiv (\xi_1, \ldots, \xi_n)$ and $\Xi \mid v \sim \textrm{Dirichlet}(1, \ldots, 1)$, as in
the weighted likelihood bootstrap of \citealp{newton:1994}.
Denoting $w_i(\xi) \equiv p_{\mathcal E}(z_i)/p_{\mathcal O}(z_i \mid x_i, v; \xi)$, the expected utility now
becomes
\begin{align}\label{equation:weightedlik}
\E _{\mathcal E}[l(\phi; V_i) \mid v ; \xi]
=\int_{v_i} l(\phi; v_i) w_i(\xi) \sum_{k=1}^n \xi_k \delta_{v_k}(v_i) \,\textrm dv_i
=\sum_{k=1}^n  w_k(\xi) \xi_k l(\phi; v_k),
\end{align}
that is, a weighted log-likelihood, motivating the estimator
$$\widehat\phi(\xi) \equiv \arg\max_\phi \sum_{k=1}^n  w_k(\xi) \xi_k l(\phi; v_k).$$
An approximate posterior distribution for $\phi$ under $\mathcal E$ can now be constructed by repeatedly sampling
the weight vectors from $\Xi \mid v \sim \textrm{Dirichlet}(1, \ldots, 1)$,
and recalculating $\widehat\phi(\xi)$ at each realization. This approach of creating a mapping
between the non-parametric specification and a parametrization relevant to inferences
is analogous to \citet{chamberlain:2003}, but adds the importance sampling weights to the Dirichlet
weights in order to make inferences across the observational and experimental regimes.

The weights $w_i(\xi)$ function as importance sampling weights in Monte Carlo integration;
they add variability to the estimation, but in the present context provide some protection towards misspecification
of the outcome model, in the sense of Section \ref{section:iptwor}.
The above did not yet address how to calculate these weights; in principle these are fully determined
by the current realization of $\xi$ under the non-parametric specification, but in practice parametric model
specifications are needed for smoothing purposes, and we need a way to link the $\xi$ and the treatment assignment
model parameters $\gamma$. For this purpose  $\gamma$ itself can be estimated through the weighted
likelihood bootstrap since this readily gives the deterministic function linking the two parametrizations;
thus in \eqref{equation:weightedlik} we choose $w_i(\xi) = p_{\mathcal E}(z_i)/p_{\mathcal O}\{z_i \mid b_i;
\widehat\gamma(\xi)\}$,
where $\widehat\gamma(\xi) \equiv \arg\max_{\gamma} \sum_{k=1}^n  \xi_k \log p(z_k \mid b_k; \gamma)$.
The probabilities $p_{\mathcal E}(z_i)$ are given by the chosen regime $\mathcal E$ that is the object of
inference;
in practice the estimation is most efficient when we choose the target regime to be as close as possible
to the observed regime $\mathcal O$; this can be achieved by fixing $p_{\mathcal E}(z_i)$ to the marginal treatment
assignment probabilities under $\mathcal O$, which would result in the usual kind of stabilized \IPT weights used
in marginal structural modelling \citep{robins:2000,hernan:2001,cole:2008}.

The marginal causal contrast may now be estimated through the expectations
\begin{align}
\E_{\Xi \mid V}\{\E_{\mathcal E}(Y_i \mid Z_i = a, v; \xi)\} &= \int_{\xi} \int_{s_i}
m\{a, s_i; \widehat\phi(\xi)\} \sum_{k=1}^n \xi_k \delta_{s_k}(s_i) p_{\mathcal O}(\xi \mid v) \d s_i \d \xi
\nonumber \\
&= \int_{\xi} \sum_{k=1}^n \xi_k m\{a, s_k; \widehat\phi(\xi)\} p_{\mathcal O}(\xi \mid v) \d \xi
\label{equation:isestimator},
\end{align}
where we used the non-parametric specification $p(s_i \mid v; \xi) = \sum_{k=1}^n \xi_k \delta_{s_k}(s_i)$, and
where again $p_{\mathcal O}(\xi \mid v)$ is replaced with the uniform Dirichlet distribution.
Since all uncertainty is contained in the parameter vector $\xi$, a posterior distribution
for the predictive mean or mean difference can be constructed as before through resampling.
The point estimator given by \eqref{equation:isestimator} is the direct Bayesian
analogue of \eqref{equation:standardization}, where the outcome model was estimated
using \IPT weighted regression. In fact, if we fix $\xi_k = 1/n$, $k = 1, \ldots, n$,
instead of considering these as unknown parameters, the two estimators are equivalent.
Thus, we conjecture that the estimator given by \eqref{equation:isestimator} has a
similar `double robustness' property as \eqref{equation:standardization}. We demonstrate this
through simulations in Section \ref{section:simulation}, but before that, we show how the semi-parametric
doubly robust estimator \eqref{eq:BangRobins} can be motivated as a posterior predictive expectation.

\subsection{Doubly robust estimation}\label{section:isdr}

In Supplementary Appendix 4 we show that under the non-parametric specification in terms of $\xi$,
the posterior predictive causal contrast may be written as
\begin{align}\label{eq:bayesdr}
\MoveEqLeft \E_{\mathcal E}(Y_i \mid Z_i = 1, v; \xi) - \E_{\mathcal E}(Y_i \mid Z_i = 0, v; \xi) \nonumber \\
&= \sum_{k=1}^n \xi_k \{y_i - m(z_k, x_k; \xi)\} \left\{\frac{z_k} {\P_{\mathcal O}(Z_k = 1 \mid x_k, v; \xi)} -
\frac{1 - z_k}{\P_{\mathcal O}(Z_k = 0 \mid x_k, v; \xi)}\right\} \nonumber \\
&\quad+ \sum_{k=1}^n \xi_k \left\{m(1, x_k; \xi) - m(0, x_k; \xi)\right\},
\end{align}
which corresponds to formulation \eqref{eq:BangRobins}. Since the non-parametric
specification places no restrictions on the conditional distributions, in practice,
to obtain an estimator, the non-parametrically specified quantities
$m(z_k, x_k; \xi)$ and $\P_{\mathcal O}(Z_k = a \mid x_k, v; \xi)$
would have to be replaced with the parametric versions
$m\{z_k, s_k; \widehat\phi(\xi)\}$ and $\P_{\mathcal O}\{Z_k = a \mid b_k; \widehat\gamma(\xi)\}$,
estimated using the weighted likelihood bootstrap, as in the previous section.
It is straightforward to see that if the outcome model is correctly specified, the expression
\eqref{eq:bayesdr} reduces to the model-based estimator (the second additive term).
This again reflects the fact that if we believe in our outcome model, the treatment assignment
model does not play a part in the inferences. However, a Bayesian might want to use an estimator
of the form \eqref{eq:bayesdr} if being restricted by two parametric constraints,
in terms of $\phi$ and $\gamma$, but not knowing which one of these is correct.
If either one of the parametric specifications is correct, \eqref{eq:bayesdr} will still reduce to the
posterior predictive mean difference, the natural Bayesian estimator. We summarize this
property in the following theorem (a proof in Supplementary Appendix 2).

\begin{theorem}\label{drproperty}
The estimator obtained by substituting the parametric specifications
$m\{z_k, s_k; \widehat\phi(\xi)\}$ and $\P_{\mathcal O}\{Z_k = a \mid b_k; \widehat\gamma(\xi)\}$
into expression \eqref{eq:bayesdr} is equivalent to the posterior predictive mean difference
if either one of these models is correctly specified.
\end{theorem}

A posterior distribution for the mean difference can be generated as in the previous section
through resampling of the parameter vectors $\xi$ and recalculating
\eqref{eq:bayesdr} at each realization; we will illustrate this in the following section.

\section{Simulation study}\label{section:simulation}

Above we have made a distinction between model misspecification due to omission of relevant
covariates, and misspecification of the parametric functional relationship
between the outcome and the covariates, and noted that all the estimators discussed in Section
\ref{section:frequentist} should be `doubly robust' against the former type of misspecification.
However, in practice the consequences of these two types of misspecification will often be similar;
they result in residual confounding. Therefore, herein we investigate how the different estimators perform when
the covariate
sets
$S_i$ and $B_i$ are not only created by a partitioning, but also a transformation of the $x_i$'s.
For this purpose, we simulated $X_{ij} \sim \mathrm{N}(0,1)$, $j = 1, \ldots, 4$,
and transformed these as $c_{ij} = |x_{ij}|/
({1-2/\pi})^{1/2}$. The true treatment assignment and outcome mechanisms were specified as
$Z_i \mid x_i \sim \mathrm{Bernoulli}(\textrm{expit}\{0.4 c_{i1} + 0.4 x_{i2} + 0.8 x_{i3}\})$ and
$Y_i \mid z_i, x_i \sim \mathrm{N}(z_i - c_{i1} - x_{i2} - x_{i4}, 1)$,
respectively. For estimation, we considered two scenarios: (I) $s_i \equiv (x_{i1}, x_{i2}, x_{i3})$
and $b_i \equiv (c_{i1}, x_{i2}, x_{i4})$ (misspecified outcome model and correctly specified treatment assignment
model),
and (II) $s_i \equiv (c_{i1}, x_{i2}, x_{i3})$ and $b_i \equiv (x_{i1}, x_{i2}, x_{i4})$
(correctly specified outcome model and misspecified treatment assignment model).

We are interested in the marginal causal contrast $\E(\mathbf Y_{i1}) - \E(\mathbf Y_{i0}) = 1$.
To estimate this, we applied the Bayesian estimators discussed in Sections \ref{section:unknown},
\ref{section:jointestimation}, and \ref{section:iptw}.
In the two-step estimation we attempted both forward sampling from the posterior distributions,
and the variance decomposition formula \eqref{equation:miformula}. In the former,
instead of Markov chain Monte Carlo, we applied normal approximations for the posterior distributions,
of the form $\Phi \mid v; \gamma \sim \mathrm{N}\{\widehat\phi(\gamma), S\}$,
where $\widehat\phi(\gamma)$ is the maximum likelihood estimate and $S$ its estimated variance-covariance matrix.
The posterior
distribution $\Gamma \mid x, z$ was approximated
using the weighted likelihood bootstrap. In the joint estimation, we again used a normal approximation,
centered at the joint maximum likelihood estimates $(\widehat\phi, \widehat\gamma)$, and the variance-covariance
matrix given by the
inverse
of the observed information at the maximum likelihood point.
In both two-step and joint estimation, the fitted models were specified as
$m\{z_i,e(b_i; \gamma), s_i; \phi\} = \phi_0 + \phi_1 z_i + \phi_2^\T s_i + \phi_3^\T g\{e(b_{i}; \gamma)\}$,
where $g$ is a cubic polynomial basis, and $e(b_{i}; \gamma) = \textrm{expit}(\gamma_0 + \gamma_1^\T b_i)$.
In the importance sampling (IS) estimator proposed in Section \ref{section:isor},
and in the importance sampling/doubly robust estimator (IS/DR) of Section \ref{section:isdr}, the fitted
treatment assignment model was the same, with the outcome model specified through
$m(z_i,s_i;\phi) = \phi_0 + \phi_1 z_i + \phi_2^\T s_i$.

For comparison to the Bayesian estimators, we also calculate naive unadjusted comparison (naive),
outcome regression adjusted for covariates $s_i$ (adjusted), the standard \IPT weighted estimator \eqref{eq:iptw}
(IPTW),
the semi-parametric doubly robust estimator \eqref{eq:BangRobins} (DR), the clever covariate version of this (CC),
the \IPT weighted outcome regression based estimator \eqref{equation:standardization} (OR/IPTW),
as well as propensity score adjusted outcome regression based estimator \eqref{equation:psadjustment} (OR/PS).
For IPTW, DR, CC, and OR/IPTW, the standard errors were estimated through
the standard frequentist nonparametric bootstrap \citep{efron:1979}. For OR/PS, to demonstrate
the variance estimation issues discussed in Section \ref{section:unknown}, we calculated both observed information based
standard errors, and the adjusted sandwich type standard errors discussed in Supplementary Appendix 3.

The results over 1000 replications are shown in Table \ref{table:simresults}.
Under scenario (I), all estimators except naive and adjusted can correct for confounding,
although the joint estimation approach produces a slight bias. In terms of efficiency,
the estimators based on propensity score adjusted outcome regression are the best, with the \IPT weighting based
estimators losing slightly. As discussed in Section \ref{section:cc}, the clever covariate approach results
in higher variability compared to the other doubly adjusted estimators.
In terms of variance estimation, the comparison between the unadjusted
and adjusted standard errors for OR/PS suggests that under this simulation setting estimation of the propensity
scores substantially reduces the variance, and not adjusting for this results in overcoverage.
The resampling based variance estimators adjust for this automatically. However, the two-step approach
to variance estimation performs poorly; as demonstrated in Supplementary Appendix 3, the two-step point estimator
has the same asymptotic variance as the other OR/PS estimators, but the two-step variance estimators unnecessarily
add a further variance component. Thus, the simulation results support the discussion in Sections
\ref{section:unknown} and \ref{section:jointestimation}; the two-step and joint estimation
approaches are difficult to justify from Bayesian arguments, and do not seem to provide practical
advantages in terms of their frequency-based properties. On the other hand, the importance sampling based Bayesian estimators
produce very similar results to the OR/IPTW and DR estimators, respectively.

Under scenario (II), all the estimators except IPTW are unbiased, which is expected
based on their previously discussed theoretical properties. When the outcome model is
correctly specified, there is also very little difference in the efficiency of the
various estimators.

\begin{table}[!ht] \small
\def~{\hphantom{0}}
\caption{Estimates for the marginal causal contrast (true value $=1$) over 1000 simulation rounds
}{%
\begin{tabular}{llcrcccr}
\multirow{2}{*}{Scenario} & \multirow{2}{*}{Estimator} & \multicolumn{1}{c}{Point}    &
\multicolumn{1}{c}{Relative} &
\multirow{2}{*}{SD} & \multirow{2}{*}{SE} & \multicolumn{1}{c}{MC} & \multicolumn{1}{c}{Coverage} \\
                          &                            & \multicolumn{1}{c}{estimate} & \multicolumn{1}{c}{bias
(\%)}  &
             &                     & \multicolumn{1}{c}{error} & \multicolumn{1}{c}{(\%)} \\
(I)    & Naive & 0.347 & $-$65.3 & 0.128 & 0.128 & 0.004 & 0.3 \\
 & Adjusted & 0.667 & $-$33.3 & 0.091 & 0.092 & 0.003 & 3.6 \\
 & IPTW & 1.001 & 0.1 & 0.135 & 0.138 & 0.005 & 94.4 \\
 & OR/PS (obs. information) & 0.997 & $-$0.3 & 0.071 & 0.095 & 0.002 & 99.3 \\
 & OR/PS (adj. sandwich) & 0.997 & $-$0.3 & 0.071 & 0.073 & 0.002 & 95.3 \\
 & DR & 0.998 & $-$0.2 & 0.087 & 0.088 & 0.003 & 94.2 \\
 & Clever covariate & 1.026 & 2.6 & 0.110 & 0.110 & 0.003 & 93.0 \\
 & OR/IPTW & 0.990 & $-$1.0 & 0.082 & 0.083 & 0.003 & 93.9 \\
 & Two-step (forward sampling) & 0.999 & $-$0.1 & 0.071 & 0.113 & 0.002 & 99.8 \\
 & Two-step (variance decomposition) & 0.995 & $-$0.5 & 0.071 & 0.112 & 0.002 & 99.8 \\
 & Joint estimation & 1.046 & 4.6 & 0.071 & 0.071 & 0.002 & 89.5 \\
 & Importance sampling & 0.991 & $-$0.9 & 0.083 & 0.080 & 0.003 & 93.6 \\
 & Importance sampling/DR & 0.997 & $-$0.3 & 0.087 & 0.086 & 0.003 & 93.9 \\
(II)   & Naive & 0.347 & $-$65.3 & 0.128 & 0.128 & 0.004 & 0.3 \\
 & Adjusted & 0.997 & $-$0.3 & 0.065 & 0.067 & 0.002 & 95.3 \\
 & IPTW & 0.629 & $-$37.1 & 0.133 & 0.131 & 0.005 & 20.3 \\
 & OR/PS (obs. information) & 0.997 & $-$0.3 & 0.071 & 0.071 & 0.002 & 95.4 \\
 & OR/PS (adj. sandwich) & 0.997 & $-$0.3 & 0.071 & 0.071 & 0.002 & 95.4 \\
 & DR & 0.999 & $-$0.1 & 0.074 & 0.074 & 0.003 & 95.9 \\
 & Clever covariate & 0.999 & $-$0.1 & 0.075 & 0.075 & 0.003 & 95.9 \\
 & OR/IPTW & 0.999 & $-$0.1 & 0.074 & 0.073 & 0.003 & 95.7 \\
 & Two-step (forward sampling) & 1.000 & 0.0 & 0.071 & 0.072 & 0.002 & 96.5 \\
 & Two-step (variance decomposition) & 0.997 & $-$0.3 & 0.071 & 0.071 & 0.002 & 95.5 \\
 & Joint estimation & 0.997 & $-$0.3 & 0.071 & 0.071 & 0.002 & 95.4 \\
 & Importance sampling & 0.999 & $-$0.1 & 0.074 & 0.072 & 0.003 & 95.2 \\
 & Importance sampling/DR & 0.999 & $-$0.1 & 0.074 & 0.073 & 0.003 & 95.5 \\
\end{tabular}}
\begin{tabnote}
The columns correspond to estimator, mean point estimate, relative bias,
Monte Carlo standard deviation (SD), mean standard error estimate (SE),
Monte Carlo (MC) error (batch means) of the mean point estimate, and 95\% confidence interval coverage probability.
The two scenarios correspond to (I) misspecified outcome model and correctly specified treatment assignment model,
and (II) correctly specified outcome model and misspecified treatment assignment model.
\end{tabnote}
\label{table:simresults}
\end{table}

\section{Discussion}\label{section:discussion}

In this paper we reviewed previously proposed Bayesian approaches for propensity score adjusted inferences,
focusing on the assumptions concerning correct model specifications. Here it
is important to make a distinction between misspecification due to omission of relevant covariates from the
outcome model,
and misspecification of the functional form of the dependency between the covariates and the outcome.
The frequentist propensity score adjusted outcome regression is robust against the former type of model
misspecification,
but this property is lost in Bayesian estimation, if the misspecified outcome model is allowed to feed back
to the estimation of the propensity scores. While feedback issue has been well documented in the literature
\citep[e.g.][]{mccandless:2009(2),zigler:2013}, and the reasons behind this were already stated by
\citet{robins:1997},
here we attempted to make the assumptions underlying the Bayesian propensity score approach more explicit. On the
other hand,
we point out that cutting this feedback in a two-step Bayesian estimation procedure unnecessarily inflates the
posterior variance estimates.

As reaching double robustness through Bayesian propensity score adjustment looks difficult, 
herein we attempted a completely different approach through posterior predictive inferences.
Our proposed approach decouples the outcome regression and treatment assignment model through
introducing the \IPT weights as importance sampling weights in Monte Carlo integration in evaluating
posterior predictive expectations. A similar procedure was used in a marginal structural modelling context by
\citet{saarela:2015a}, improved to its present form in \citet{saarela:2015b}.
While they used the importance sampling approach for estimating marginal outcome models in a longitudinal setting,
herein we showed that in a point treatment setting the combination of importance sampling
and posterior predictive inferences can be used to motivate weighted outcome regression or semi-parametric doubly robust
inferences.
Such a possibility was mentioned, but not formally justified,
by \citet{saarela:2015c} who applied the importance sampling procedure in the context of estimating optimal treatment regimes.
The disadvantage of the importance sampling approach is the same as in the corresponding frequentist
\IPT weighted inference procedures: the importance sampling weights add variability to the point estimator.
In order to control this, a standard approach would be to truncate the weights \citep[e.g.][]{xiao:2013},
which would also be possible in the importance sampling context \citep{ionides:2008}. Recently,
Vehtari \& Gelman (2015, \url{arXiv:1507.02646v2}) suggested probabilistic truncation of importance sampling
weights; studying this possibility in the present context is a topic for further research.

\section*{Acknowledgement}

The authors acknowledge support of the Natural Sciences and Engineering Research Council (NSERC) of Canada.
\appendix
\section*{Supplementary material}
Supplementary material available includes Supplementary Appendices 1-4 referred to herein,
containing proofs to theorems and other technical material.

\section*{Appendix 1}\label{section:otoe}

\subsection*{Causal inference as a prediction problem}

The estimator (2) can be motivated without the use of potential outcomes notation
as a posterior predictive expectation for a new observation under a hypothetical completely randomized setting 
$\mathcal E$ where 
$Z_i {\perp\!\!\!\!\perp_{\mathcal E}} X_i$ and the probabilities $\P_{\mathcal E}(Z_i = a)$ are known constants
\citep[cf. the randomized trial measure discussed by][]{roysland:2011}. The data are observed under a setting 
$\mathcal O$, where
$Z_i {\notindep_{\mathcal O}} X_i$, and causal inference then corresponds to inference
across these regimes. We can now write for $i \notin \{1, \ldots, n\}$
\begin{align}
\nonumber \MoveEqLeft\E_{\mathcal E}(Y_i \mid Z_i = a, v)
\\&= 
\frac{\int_{\phi, \psi} \int_{x_i} \int_{y_i}
y_i p(y_i \mid Z_i = a, x_i; \phi) \P_{\mathcal E}(Z_i = a) p(x_i ; \psi) \pi_n(\phi)\pi_n(\psi) \d y_i \d 
x_i \d \phi \d \psi}
{\int_{\phi, \psi} \int_{x_i} \int_{y_i} p(y_i \mid Z_i = a, x_i; \phi) \P_{\mathcal E}(Z_i = a) p(x_i; \psi) 
\pi_n(\phi)\pi_n(\psi) \d y_i \d x_i \d \phi \d \psi} \nonumber \\
&= \int_{\phi, \psi} \int_{x_i}
m(a, x_k; \phi) p(x_i ; \psi) \pi_n(\phi)\pi_n(\psi)\d x_i \d \phi \d \psi \label{equation:ppredictive1} 
\\
&= \int_{\phi} \frac{1}{n} \sum_{k=1}^n 
m(a, x_k; \phi) \pi_n(\phi)\d \phi \label{equation:ppredictive2},
\end{align}
where the last form was obtained by choosing the non-parametric specification
${\int_{\psi} p(x_i ; \psi) \pi_n(\psi)\d \psi} = {p(x_i \mid x_1, \ldots, x_n)} = {\sum_{k=1}^n 
\delta_{x_k}(x_i)/n}$. Alternatively, in \eqref{equation:ppredictive1} one could use the Bayesian bootstrap 
\citep{rubin:1981} specification
$p(x_i \mid x_1, \ldots, x_n; \xi) = \sum_{k=1}^n \xi_k \delta_{x_k}(x_i)$,
where $\xi = (\xi_1, \ldots, \xi_n)$, with $\pi_n(\xi)$ taken to be
a uniform Dirichlet distribution (see Section 6).
Obtaining \eqref{equation:ppredictive2} also required assuming that
$p_{\mathcal E}(y_i \mid z_i, x_i; \phi) = p_{\mathcal O}(y_i \mid z_i, x_i; \phi)
\equiv p(y_i \mid z_i, x_i; \phi)$ and
and $p_{\mathcal E}(x_i ; \psi) = p_{\mathcal O}(x_i ; \psi) \equiv p(x_i ; \psi)$,
which corresponds to the stability assumption discussed by \citet{dawid:2010}.

\section*{Appendix 2}\label{section:proofs}
\subsection*{Proofs to theorems}

\begin{proof}[to Theorem 1]
If (i) holds true, then also 
$(\mathbf Y_{0i}, \mathbf Y_{1i}) \independent e(B_i) \mid S_i$, and the propensity
score adjustment does not add information. If on the other hand (ii) holds true,
$\{e(B_i), S_i\}$ has jointly the balancing property $Z_i \independent X_i \mid \{e(B_i), S_i\}$. 
This follows from Theorem 2 of \citet[][p. 44]{rosenbaum:1983} and also implies
that $(\mathbf Y_{0i}, \mathbf Y_{1i}) \independent Z_i \mid \{e(B_i), S_i\}$
\citep[][Theorem 3]{rosenbaum:1983}. Now
\begin{align*}
\E\{Y_i \mid Z_i, e(B_i), S_i\} & = \E\{(1 - Z_i) \mathbf Y_{0i} \mid Z_i, e(B_i), S_i\}
    + \E\{Z_i \mathbf Y_{1i} \mid Z_i, e(B_i), S_i\} \\
&= (1 - Z_i)\E\{\mathbf Y_{0i} \mid Z_i, e(B_i), S_i\} + Z_i \E\{\mathbf Y_{1i} \mid Z_i, e(B_i), S_i\} \\
&= (1 - Z_i)\E\{\mathbf Y_{0i} \mid e(B_i), S_i\} + Z_i \E\{\mathbf Y_{1i} \mid e(B_i), S_i\},
\end{align*}
and further,
\begin{align}
\MoveEqLeft\int_{x_i} \E\{Y_i \mid Z_i = 1, e(b_i), s_i\} p(x_i)\d x_i - \int_{x_i} \E\{Y_i \mid Z_i = 0, e(b_i), 
s_i\} 
p(x_i)\d x_i\label{equation:causalcontrast}\\
& \qquad = \int_{x_i} \E\{\mathbf Y_{1i} \mid e(b_i), s_i\} p(x_i)\d x_i - \int_{x_i} \E\{\mathbf Y_{0i} \mid 
e(b_i), s_i\} p(x_i)\d x_i \nonumber \\
& \qquad = \E(\mathbf Y_{1i}) - \E(\mathbf Y_{0i}).\nonumber
\end{align}
The consistency of the estimator (3) then relies on being
able to consistently estimate the quantities in \eqref{equation:causalcontrast}.
\end{proof}

\begin{proof}[to Theorem 2]
Consider first the expectation of (8) under the assumption that 
(i) holds true. Now 
\begin{align*}
&\E\left\{ \mathbf 1_{\{z_i = a\}}\frac{\textstyle \frac{\partial}{\partial \phi}\log p(\mathbf y_{ai} \mid s_i, 
\phi)}{\P(Z_i = a \mid b_i)} \right\} \\
&= \int_{x_i}\int_{\mathbf y_{ai}} 
\sum_{z_i} \mathbf 1_{\{z_i = a\}}\frac{\textstyle \frac{\partial}{\partial \phi}\log p(\mathbf y_{ai} \mid s_i, 
\phi)}{\P(Z_i = a \mid b_i)}
f(\mathbf y_{ai} \mid z_i, x_i)f(z_i \mid x_i)f(x_i) \,\textrm d\mathbf y_{ai} \,\textrm dx_i\\
&= \int_{x_i}\left\{ \int_{\mathbf y_{ai}} \frac{\textstyle \frac{\partial}{\partial \phi}p(\mathbf y_{ai} \mid 
s_i; \phi)}{p(\mathbf y_{ai} \mid s_i; \phi)}
f(\mathbf y_{ai} \mid x_i) \,\textrm d\mathbf y_{ai}\right\}\frac{\P(Z_i = a \mid x_i)}{\P(Z_i = a \mid 
b_i)}f(x_i)\,\textrm dx_i \\
&= \int_{x_i}\left\{\frac{\textstyle \partial}{\partial \phi}\int_{\mathbf y_{ai}}  p(\mathbf y_{ai} \mid s_i; 
\phi) 
\,\textrm d\mathbf y_{ai}\right\}\frac{\P(Z_i = a \mid x_i)}{\P(Z_i = a \mid b_i)}f(x_i)\,\textrm dx_i = 0,
\end{align*}
which followed because now $p(\mathbf y_{ai} \mid s_i; \phi) = f(\mathbf y_{ai} \mid x_i)$ at the true parameter 
value.
Thus, the misspecified weights do not influence the estimation (in terms of bias) 
as long as the outcome model is correctly specified.

Under the assumption that (ii) holds true, we have in turn that
\begin{align*}
&\E\left\{ \mathbf 1_{\{z_i = a\}}\frac{\textstyle \frac{\partial}{\partial \phi}\log p(\mathbf y_{ai} \mid s_i; 
\phi)}{\P(Z_i = a \mid b_i)} \right\} \\
&= \int_{x_i}\int_{\mathbf y_{ai}} 
\sum_{z_i} \mathbf 1_{\{z_i = a\}}\frac{\textstyle \frac{\partial}{\partial \phi}\log p(\mathbf y_{ai} \mid s_i; 
\phi)}{\P(Z_i = a \mid b_i)}
f(\mathbf y_{ai} \mid z_i, x_i)f(z_i \mid x_i)f(x_i) \,\textrm d\mathbf y_{ai} \,\textrm dx_i\\
&= \int_{x_i}\int_{\mathbf y_{ai}} 
\frac{\partial}{\partial \phi}\log p(\mathbf y_{ai} \mid s_i; \phi)
f(\mathbf y_{ai} \mid x_i)f(x_i) \,\textrm d\mathbf y_{ai} \,\textrm dx_i,
\end{align*}
since now $p(z_i \mid b_i) = f(z_i \mid x_i)$.
Using the partitioning $x_i = (s_i, r_i)$, we can write the last form in above as
\begin{align*}
&\int_{x_i}\int_{\mathbf y_{ai}} 
\frac{\partial}{\partial \phi}\log p(\mathbf y_{ai} \mid s_i; \phi)
f(\mathbf y_{ai} \mid x_i)f(x_i) \,\textrm d\mathbf y_{ai} \,\textrm dx_i \\
&= \int_{s_i}\int_{r_i}\int_{\mathbf y_{ai}} 
\frac{\partial}{\partial \phi}\log p(\mathbf y_{ai} \mid s_i; \phi)
f(\mathbf y_{ai}, s_i, r_i) \,\textrm d\mathbf y_{ai} \,\textrm ds_i \,\textrm dr_i \\
&= \int_{s_i}\int_{\mathbf y_{ai}} 
\frac{\frac{\partial}{\partial \phi} p(\mathbf y_{ai} \mid s_i; \phi)}{p(\mathbf y_{ai} \mid s_i; \phi)}
f(\mathbf y_{ai} \mid s_i) f(s_i) \,\textrm d\mathbf y_{ai} \,\textrm ds_i \\
&= \int_{s_i}
\left\{\frac{\partial}{\partial \phi} \int_{\mathbf y_{ai}} p(\mathbf y_{ai} \mid s_i; \phi)
\,\textrm d\mathbf y_{ai} \right\} f(s_i) \,\textrm ds_i = 0.
\end{align*}
Thus, even though the outcome regression does not include a sufficient set of confounders, 
through the IPT weighting we can still obtain valid estimates for 
the conditional distributions $p(\mathbf y_{ai} \mid s_i; \phi)$.
\end{proof}

\begin{proof}[to theorem 3]
Now the marginal posterior distribution of the parameters $\gamma$ becomes
\begin{align*}
p(\gamma \mid v) &= \int_{\phi, \psi} p(\phi, \gamma, \psi \mid v) \d \phi \d \psi\\
&\propto \int_{\phi, \psi} \left\{\prod_{i=1}^n
p(y_i \mid z_i, x_i; \phi) p(z_i \mid x_i; \gamma) p(x_i ; \psi) \right\} \pi_0(\phi)\pi_0(\psi)\pi_0(\gamma) 
\d\phi\d\psi\d\gamma
\\
&\propto \prod_{i=1}^n p(z_i \mid x_i; \gamma) \pi_0(\gamma) \\
&\propto p(\gamma \mid x, z).
\end{align*}
\end{proof}

\begin{proof}[to Theorem 4]

The estimator obtained through substituting in the parametric models is
\begin{align}\label{eq:bayesdrestimator}
&\sum_{k=1}^n \xi_k \left[y_i - m\{z_k, s_k; \widehat\phi(\xi)\}\right] \left[\frac{z_k} {\P_{\mathcal O}\{Z_k = 1 \mid b_k; 
\widehat\gamma(\xi)\}} - 
\frac{1 - z_k}{\P_{\mathcal O}\{Z_k = 0 \mid b_k; \widehat\gamma(\xi)\}}\right] \nonumber \\
&\quad+ \sum_{k=1}^n \xi_k \left[m\{1, s_k; \widehat\phi(\xi)\} - m\{0, s_k; \widehat\phi(\xi)\}\right].
\end{align}
First, if the outcome model is correctly specified in the sense that 
$m\{z_k, s_k; \widehat\phi(\xi)\} = m(z_k, x_k; \xi)$, we get 
\begin{align*}
\MoveEqLeft \P_{\mathcal E}(Z_k = a) \sum_{k=1}^n \xi_k  \frac{\mathbf 1_{\{z_k = a\}} m\{z_k, s_k; \widehat\phi(\xi)\}}
{\P_{\mathcal O}\{Z_k = a \mid b_k; \widehat\gamma(\xi)\}} \\
&= \sum_{k=1}^n \xi_k \mathbf 1_{\{z_k = a\}} m\{z_k, s_k; \widehat\phi(\xi)\} \frac{p_{\mathcal E}(z_k)}
{p_{\mathcal O}\{z_k \mid b_k; \widehat\gamma(\xi)\}} \\
&= \int_{z_i, x_i} \mathbf 1_{\{z_i = a\}} m\{z_i, s_i; \widehat\phi(\xi)\} \frac{p_{\mathcal E}(z_i)}
{p_{\mathcal O}\{z_i \mid b_i; \widehat\gamma(\xi)\}} \sum_{k=1}^n \xi_k \delta_{(z_k, x_k)}(z_i, x_i) \d z_i \d x_i \\
&= \int_{y_i, z_i, x_i} \mathbf 1_{\{z_i = a\}} y_i \frac{p_{\mathcal E}(z_i)}
{p_{\mathcal O}\{z_i \mid b_i; \widehat\gamma(\xi)\}} p_{\mathcal O}(y_i \mid z_i, x_i, v; \xi)
p_{\mathcal O}(z_i, x_i \mid v; \xi) \d y_i \d z_i \d x_i\\
&= \int_{v_i} \mathbf 1_{\{z_i = a\}} y_i \frac{p_{\mathcal E}(z_i)}
{p_{\mathcal O}\{z_i \mid b_i; \widehat\gamma(\xi)\}}
p_{\mathcal O}(v_i \mid v; \xi) \d v_i\\
&= \P_{\mathcal E}(Z_k = a) \sum_{k=1}^n \xi_k \frac{\mathbf 1_{\{z_k = a\}} y_k}
{\P_{\mathcal O}\{Z_k = a \mid b_k; \widehat\gamma(\xi)\}},
\end{align*}
because the second to last form is equivalent to \eqref{equation:e1} in Supplementary Appendix 4. Thus, the first
summation term in \eqref{eq:bayesdrestimator} cancels out, leaving only the model based estimator,
which itself is now equivalent to the posterior predictive mean difference.

On the other hand, if the treatment assignment model is correctly specified in the
sense that \\$\P_{\mathcal O}\{Z_k = a \mid b_k; \widehat\gamma(\xi)\} = \P_{\mathcal O}(Z_k = a \mid x_k, v; \xi)$, we get
\begin{align*}
\MoveEqLeft \P_{\mathcal E}(Z_k = a) \sum_{k=1}^n \xi_k  \frac{\mathbf 1_{\{z_k = a\}} m\{z_k, s_k; \widehat\phi(\xi)\}}
{\P_{\mathcal O}\{Z_k = a \mid b_k; \widehat\gamma(\xi)\}} \\
&= \sum_{k=1}^n \xi_k \mathbf 1_{\{z_k = a\}} m\{z_k, s_k; \widehat\phi(\xi)\} \frac{p_{\mathcal E}(z_k)}
{p_{\mathcal O}\{z_k \mid b_k; \widehat\gamma(\xi)\}} \\
&= \int_{z_i, x_i} \mathbf 1_{\{z_i = a\}} m\{z_i, s_i; \widehat\phi(\xi)\} \frac{p_{\mathcal E}(z_i)}
{p_{\mathcal O}\{z_i \mid b_i; \widehat\gamma(\xi)\}} \sum_{k=1}^n \xi_k \delta_{(z_k, x_k)}(z_i, x_i) \d z_i \d x_i \\
&= \int_{z_i, x_i} \mathbf 1_{\{z_i = a\}} m\{z_i, s_i; \widehat\phi(\xi)\} \frac{p_{\mathcal E}(z_i)}
{p_{\mathcal O}\{z_i \mid b_i; \widehat\gamma(\xi)\}} p_{\mathcal O}(z_i \mid x_i, v; \xi) p_{\mathcal O}(x_i \mid v; \xi) \d z_i 
\d x_i \\
&= \int_{z_i, x_i} \mathbf 1_{\{z_i = a\}} m\{z_i, s_i; \widehat\phi(\xi)\} p_{\mathcal E}(z_i)
p_{\mathcal O}(x_i \mid v; \xi) \d z_i \d x_i \\
&= \P_{\mathcal E}(Z_i = a) \int_{x_i} m\{a, s_i; \widehat\phi(\xi)\}
\sum_{k=1}^n \xi_k \delta_{(x_k)}(x_i) \d x_i \\
&= \P_{\mathcal E}(Z_k = a) \sum_{k=1}^n 
 \xi_k m\{a, s_k; \widehat\phi(\xi)\}.
\end{align*}
Therefore, the estimator \eqref{eq:bayesdrestimator} now reduces to 
\begin{align*}
&\sum_{k=1}^n \xi_k y_i \left[\frac{z_k} {\P_{\mathcal O}\{Z_k = 1 \mid b_k; \widehat\gamma(\xi)\}} - 
\frac{1 - z_k}{\P_{\mathcal O}\{Z_k = 0 \mid b_k; \widehat\gamma(\xi)\}}\right],
\end{align*}
which is again equivalent to the posterior predictive mean difference (see Supplementary Appendix 4).
\end{proof}

\section*{Appendix 3}\label{section:appendix}
\subsection*{On the frequency-based properties of the two-step approach}

Trivially, if the outcome model is correctly specified, 
then $\Phi \independent \Gamma \mid V$ and (9) reduces to \eqref{equation:ppredictive2}. 
The interesting situations are naturally those where this is not the case.
We denote the log-likelihood by $q_i(\phi; \gamma) =  \log p\{y_{i} \mid z_{i}, e(b_{i}; \gamma), s_{i}; \phi)\}$ and
$q(\phi; \gamma) \equiv \sum_{i=1}^n q_i(\phi; \gamma)$ and consider
the quasi-maximum likelihood estimator $\widehat \phi(\widehat \gamma) \equiv \arg \max_{\phi} q(\phi; \widehat 
\gamma)$.
If the treatment assignment model is correctly specified, $\widehat \gamma \rightarrow 
\gamma_0$.
In addition, we assume that with any fixed value of $\gamma$, $\widehat \phi(\gamma) \rightarrow \phi_0(\gamma)$,
where $\phi_0(\gamma)$ is the parameter vector which minimizes the Kullback-Leibler
divergence to the true outcome model \citep[e.g.][p. 4]{white:1982}.
Thus, by the law of large numbers and continuous mapping, we can write in the usual way that
\begin{equation*}
\frac{1}{n} \sum_{i = 1}^n [q_i(\phi; \widehat \gamma) - q_i\{\phi_0(\gamma_0); \gamma_0\}]
\rightarrow E\left[q_i(\phi; \gamma_0) - q_i\{\phi_0(\gamma_0); \gamma_0\} \right],
\end{equation*}
where the right hand side is maximized at zero when $\phi = \phi_0(\gamma_0)$, at which point
\[\E\{Y_i \mid Z_i = a, e(b_i; \gamma_0), s_i; \phi_0(\gamma_0)\} =
\E\{\mathbf Y_{ia} \mid e(b_i; \gamma_0), s_i; \phi_0(\gamma_0)\}.\]
Since we also have that the posterior
$p(\gamma \mid x, y) \rightarrow \delta_{\gamma_0}(\gamma)$ in distribution,
we can then conjecture that posterior predictive inferences based on 
(9) will be asymptotically uncounfounded.

With the definitions
\begin{align*}
U^{\phi}(\phi; \gamma) &\equiv \partial q(\phi; \gamma)/\partial \phi, \\
U^{\phi\phi}(\phi; \gamma) &\equiv \partial^2 q(\phi; \gamma)/\partial \phi^2, \\
U^{\phi\gamma}(\phi; \gamma) &\equiv \partial^2 q(\phi; \gamma)/\partial \phi\partial \gamma, \\
U^{\gamma}(\gamma) &\equiv \partial \textstyle \sum_{i=1}^n \log p(z_i \mid b_i; \gamma)/\partial \gamma, \\
U^{\gamma\gamma}(\gamma) &\equiv \partial^2 \textstyle \sum_{i=1}^n \log p(z_i \mid b_i; \gamma)/\partial \gamma^2,
\end{align*}
and noting that $U^{\phi}(\widehat \phi; \gamma^{(j)}) = 0$ for each $\gamma^{(j)}$, $j = 1, \ldots, m$,
we can consider the first order Taylor expansion of
$U^{\phi}(\widehat \phi; \gamma^{(j)})$ around the true parameter values $(\phi_0, \gamma_0)$, which becomes
\begin{align*}
0 &= \frac{1}{n} U^{\phi}(\widehat \phi; \gamma^{(j)}) \\
&\approx \frac{1}{n} U^{\phi}(\phi_0; \gamma_0) +
\frac{1}{n} U^{\phi\phi}(\phi_0; \gamma_0) \{\widehat \phi(\gamma^{(j)}) - \phi_0\} + \frac{1}{n} 
U^{\phi\gamma}(\phi_0; \gamma_0)(\gamma^{(j)} - \gamma_0) \\
&\approx \frac{1}{n} U^{\phi}(\phi_0; \gamma_0) + \E\{U_i^{\phi\phi}(\phi_0; \gamma_0)\} \{\widehat 
\phi(\gamma^{(j)}) - 
\phi_0\} + \E\{U_i^{\phi\gamma}(\phi_0; \gamma_0)\}(\gamma^{(j)} - \gamma_0),
\end{align*}
and further,
\begin{align*}
0 &= \frac{1}{m} \sum_{j=1}^m \frac{1}{n} U^{\phi}(\widehat{\phi}; \gamma^{(j)}) \\
&\approx \frac{1}{n} U^{\phi}(\phi_0; \gamma_0) + \E\{U_i^{\phi\phi}(\phi_0; \gamma_0)\} \Bigg\{ \frac{1}{m} 
\sum_{j=1}^m\widehat{\phi}(\gamma^{(j)}) - \phi_0\Bigg\} +\E\{U_i^{\phi\gamma}(\phi_0; 
\gamma_0)\}\left(\widehat{\gamma} - 
\gamma_0\right),
\end{align*}
if $\frac{1}{m} \sum_{j=1}^m\gamma^{(j)} \approx \widehat{\gamma}$. Hence,
\begin{align*}
\sqrt{n} \Bigg\{ \frac{1}{m} \sum_{j=1}^m\widehat \phi(\gamma^{(j)}) - \phi_0\Bigg\} 
&\approx \E\{- U_i^{\phi\phi}(\phi_0; \gamma_0)\}^{-1}\\&\quad \times \left[\frac{\sqrt{n}}{n} U^{\phi}(\phi_0; 
\gamma_0) + 
\E\{U_i^{\phi\gamma}(\phi_0; \gamma_0)\}\sqrt{n}(\widehat \gamma - \gamma_0) \right].
\end{align*}
Here we have, by another first order expansion around $\gamma_0$, that
\begin{equation*}
\sqrt{n}(\widehat \gamma - \gamma_0) \approx \E\{-U_i^{\gamma\gamma}(\gamma_0)\}^{-1} \frac{\sqrt{n}}{n} 
U^{\gamma}(\gamma_0),
\end{equation*}
so finally,
\begin{align*}
\MoveEqLeft\sqrt{n} \Bigg\{ \frac{1}{m} \sum_{j=1}^m\widehat \phi(\gamma^{(j)}) - \phi_0\Bigg\} \approx \E\{- 
U_i^{\phi\phi}(\phi_0; 
\gamma_0)\}^{-1} \\&\qquad \qquad\qquad\times\left(\frac{\sqrt{n}}{n} \sum_{i=1}^n \left[ 
U_i^{\phi}(\phi_0; \gamma_0) + \E\{U_i^{\phi\gamma}(\phi_0; \gamma_0)\} \E\{- U_i^{\gamma\gamma}(\gamma_0)\}^{-1} 
U_i^{\gamma}(\gamma_0)\right]\right).
\end{align*}
We may similarly expand $U^{\phi}(\widehat \phi; \widehat \gamma)$ where the parameters $\gamma$
have been fixed to their maximum likelihood estimates around $(\phi_0, \gamma_0)$ as
\begin{align*}
0 &= \frac{1}{n} U^{\phi}(\widehat \phi; \widehat \gamma) \\
&\approx \frac{1}{n} U^{\phi}(\phi_0; \gamma_0) + \E\{U_i^{\phi\phi}(\phi_0; \gamma_0)\} (\widehat \phi(\widehat 
\gamma) - 
\phi_0) + \E\{U_i^{\phi\gamma}(\phi_0; \gamma_0)\}(\widehat \gamma - \gamma_0)
\end{align*}
to find that
\begin{align*}
\sqrt{n} \left\{\widehat \phi(\widehat \gamma) - \phi_0\right\} &\approx \E\{- U_i^{\phi\phi}(\phi_0; 
\gamma_0)\}^{-1}
 \frac{\sqrt{n}}{n} \sum_{i=1}^n B_i(\phi_0; \gamma_0) ,
\end{align*}
where
\begin{equation*}
B_i(\phi_0; \gamma_0) \equiv U_i^{\phi}(\phi_0; \gamma_0) + \E\{U_i^{\phi\gamma}(\phi_0; \gamma_0)\} \E\{- 
U_i^{\gamma\gamma}(\gamma_0)\}^{-1} U_i^{\gamma}(\gamma_0).
\end{equation*}
Since the two estimators $\frac{1}{m} \sum_{j=1}^m\widehat \phi(\gamma^{(j)})$ and
$\widehat \phi(\widehat \gamma)$ have the same linear approximation which is a sum
of independent terms, we conclude that they have the same asymptotic distribution, 
\begin{equation*}
\sqrt{n} (\widehat \phi - \phi_0) \rightarrow \mathrm{N}\left[0, \E\{-U_i^{\phi\phi}(\phi_0; \gamma_0)\}^{-1} 
\V\{B_i(\phi_0; \gamma_0)\} \E\{-U^{\phi\phi}(\phi_0; \gamma_0)^\T \}^{-1}\right].
\end{equation*}

Fitting the regression model $y_{i} = \phi_0 + \phi_1 z_{i} + \phi_2^\T s_{i} + 
\phi_3^\T g\{e(b_{i}; \gamma)\} + \varepsilon_{1i}$ to estimate the parameter of interest $\phi_1$ is
numerically equivalent to fitting the sequence of regressions
$y_{i} = \nu^\T s_i^*(\widehat \gamma) + \varepsilon_{2i}$,
$z_{i} = \alpha^\T s_i^*(\widehat \gamma) + \varepsilon_{3i}$ and
$\{y_{i} - \widehat\nu^\T s_i^*(\widehat \gamma)\} = \phi_1 \{z_{i} - \widehat\alpha^\T s_i^*(\widehat \gamma)\} + 
\varepsilon_{4i}$,
where $s_i^*(\gamma) \equiv [s_{i}, g\{e(b_{i}; \gamma)\}]$. Denoting the estimating function
corresponding to the last regression as
\begin{equation*}
U^{\phi_1}\{\phi_1, \widehat\gamma, \widehat\nu(\widehat \gamma), \widehat\alpha(\widehat \gamma)\} \equiv 
\sum_{i=1}^n \{z_{i} - 
\widehat\alpha^\T s_i^*(\widehat \gamma)\}
\left[\{y_{i} - \widehat\nu^\T s_i^*(\widehat \gamma)\} - \phi_1 \{z_{i} - \widehat\alpha^\T s_i^*(\widehat 
\gamma)\}\right],
\end{equation*}
and the partial derivatives of this as e.g. $U^{\phi_1\gamma} \equiv \partial U^{\phi_1}/\partial \gamma$,
we can expand this around $(\phi_{10}, \gamma_0, \nu_0, \alpha_0)$, where $\nu_0 \equiv \nu_0(\gamma_0)$ and
$\alpha_0 \equiv \alpha_0(\gamma_0)$ are the limiting values of the nuisance parameters, as
\begin{align*}
\frac{1}{n} U^{\phi_1}\{\phi_1, \widehat \gamma, \widehat\nu(\widehat \gamma), \widehat\alpha(\widehat \gamma)\}
&\approx \frac{1}{n} U^{\phi_1}(\phi_{10}, \gamma_0, \nu_0, \alpha_0)
+ \E\{U_i^{\phi_1\gamma}(\phi_{10}, \gamma_0, \nu_0, \alpha_0)\} (\widehat \gamma - \gamma_0) \\
&\quad+ \E\{U_i^{\phi_1\nu}(\phi_{10}, \gamma_0, \nu_0, \alpha_0)\} (\widehat \nu - \gamma_0) \\
&\quad+ \E\{U_i^{\phi_1\alpha}(\phi_{10}, \gamma_0, \nu_0, \alpha_0)\} (\widehat \alpha - \gamma_0) \\
&= \frac{1}{n} U^{\phi_1}(\phi_{10}, \gamma_0, \nu_0, \alpha_0)
+ \E\{U_i^{\phi_1\gamma}(\phi_{10}, \gamma_0, \nu_0, \alpha_0)\} (\widehat \gamma - \gamma_0) \\
&\approx \frac{1}{n} U^{\phi_1}(\phi_1, \widehat \gamma, \nu_0, \alpha_0),
\end{align*}
since here $\E\{U_i^{\phi_1\nu}(\phi_{10}, \gamma_0, \nu_0, \alpha_0)\} = \E\{U_i^{\phi_1\alpha}(\phi_{10}, 
\gamma_0, 
\nu_0, 
\alpha_0)\} = 0$.
We can now see that the Theorem 1 of \citet[][p. 935]{henmi:2004} applies to the last
form here, implying that $\mathrm{avar}(\widehat\phi_1) \le \mathrm{avar}(\tilde\phi_1)$,
where $\widehat\phi_1$ is the solution to $U(\phi_1, \widehat \gamma, \nu_0, \alpha_0) = 0$ and
$\tilde\phi_1$ is the solution to $U(\phi_1, \gamma_0, \nu_0, \alpha_0) = 0$.

\section*{Appendix 4}\label{section:bayesdr}
\subsection*{The doubly robust estimator as a posterior predictive expectation}

We first note that because
\begin{align*}
\MoveEqLeft \int_{v_i} \mathbf 1_{\{z_i = a\}} y_i p_{\mathcal E}(v_i \mid v; \xi) \d v_i \\
&= \int_{y_i, x_i} y_i p_{\mathcal E}(y_i \mid Z_i = a, x_i, v; \xi) \P_{\mathcal E}(Z_i = a) p_{\mathcal E}(x_i 
\mid v ; \xi) \d y_i \d x_i,
\end{align*}
we have that
\begin{align*}
\E_{\mathcal E}(Y_i \mid Z_i = 1, v; \xi)
&= \E_{\mathcal E}\left\{\left.\frac{Z_i Y_i}{\P_{\mathcal E}(Z_i = 1)} \,\right|\, v; \xi\right\}
\shortintertext{and}
\E_{\mathcal E}(Y_i \mid Z_i = 0, v; \xi)
&= \E_{\mathcal E}\left\{\left.\frac{(1 - Z_i) Y_i}{\P_{\mathcal E}(Z_i = 0)} \,\right|\, v; \xi\right\}.
\end{align*}

The usual IPT-weighted estimator for a marginal causal contrast may be derived
through a posterior predictive argument as follows. First,
\begin{align}\label{equation:e1}
\E_{\mathcal E}[Z_i Y_i \mid v; \xi]
&= \int_{v_i} z_i y_i p_{\mathcal E}(v_i \mid v; \xi) \d v_i \nonumber \\
&= \int_{v_i} z_i y_i \frac{p_{\mathcal E}(v_i \mid v; \xi)}{p_{\mathcal O}(v_i \mid v; \xi)} p_{\mathcal O}(v_i \mid v; 
\xi)  \d v_i \nonumber \\
&= \int_{v_i} z_i y_i \frac{p_{\mathcal E}(y_i \mid z_i, x_i, v; \xi)p_{\mathcal E}(z_i)p_{\mathcal E}(x_i \mid v, \xi)}
{p_{\mathcal O}(y_i \mid z_i, x_i, v; \xi)p_{\mathcal O}(z_i \mid x_i, v; \xi)p_{\mathcal O}(x_i \mid v ; \xi)} p_{\mathcal O}(v_i 
\mid v; \xi) 
\d v_i \nonumber \\
&= \int_{v_i} z_i y_i \frac{p_{\mathcal E}(z_i)}
{p_{\mathcal O}(z_i \mid x_i, v; \xi)} p_{\mathcal O}(v_i \mid v; \xi) \d v_i \\
&= \int_{v_i} z_i y_i \frac{p_{\mathcal E}(z_i)}
{p_{\mathcal O}(z_i \mid x_i, v; \xi)} \sum_{k=1}^n \xi_k \delta_{v_k}(v_i) \d v_i \nonumber \\
&= \sum_{k=1}^n \xi_k z_k y_k \frac{p_{\mathcal E}(z_k)}
{p_{\mathcal O}(z_k \mid x_k, v; \xi)} \nonumber \\
&= \P_{\mathcal E}(Z_k = 1) \sum_{k=1}^n \xi_k \frac{z_k y_k}
{\P_{\mathcal O}(Z_k = 1 \mid x_k, v; \xi)},\nonumber 
\end{align}
and thus,
\begin{align*}
\E_{\mathcal E}\left\{\left.\frac{Z_i Y_i}{\P_{\mathcal E}(Z_i = 1)} \,\right|\, v; \xi\right\} &= \sum_{k=1}^n 
\xi_k \frac{z_k y_k}
{\P_{\mathcal O}(Z_k = 1 \mid x_k, v; \xi)}.
\intertext{Similarly,}\E_{\mathcal E}\left\{\left.\frac{(1 - Z_i) Y_i}{\P_{\mathcal E}(Z_i = 0)}\,\right|\, v; 
\xi\right\} &= \sum_{k=1}^n \xi_k \frac{(1 - z_k) y_k}
{\P_{\mathcal O}(Z_k = 0 \mid x_k, v; \xi)},
\end{align*}
and
\begin{align*}
\MoveEqLeft \E_{\mathcal E}\left\{\left.\frac{Z_i Y_i}{\P_{\mathcal E}(Z_i = 1)} - \frac{(1 - Z_i) 
Y_i}{\P_{\mathcal E}(Z_i = 0)} \,\right|\, v; \xi\right\} \\
&= \sum_{k=1}^n \xi_k y_k \left\{\frac{z_k} {\P_{\mathcal O}(Z_k = 1 \mid x_k, v; \xi)} - \frac{1 - z_k}{\P_{\mathcal O}(Z_k = 0 
\mid x_k, v; \xi)}\right\}.
\end{align*}

On the other hand, the usual outcome model based estimator may be motivated similarly as in Supplementary Appendix 1 through
\begin{align*}
\E_{\mathcal E}(Y_i \mid Z_i = a, v; \xi) &= 
\int_{x_i}
\left\{ \int_{y_i} y_i p_{\mathcal O}(y_i \mid Z_i = a, x_i, v; \xi) \d y_i \right\} p_{\mathcal O}(x_i \mid v; 
\xi) \d x_i\\
&= \int_{x_i} m(a, x_i; \xi) \sum_{k=1}^n \xi_k \delta_{x_k}(x_i) \d x_i\\
&= \sum_{k=1}^n \xi_k m(a, x_k; \xi),
\end{align*}
and
\begin{align*}
\E_{\mathcal E}(Y_i \mid Z_i = 1, v; \xi) - \E_{\mathcal E}(Y_i \mid Z_i = 0, v; \xi) = \sum_{k=1}^n \xi_k \left\{m(1, x_k; \xi) - 
m(0, x_k; \xi)\right\}.
\end{align*}
Finally, we note that we can write \eqref{equation:e1} alternatively as
\begin{align*}
\MoveEqLeft \E_{\mathcal E}(Z_i Y_i \mid v; \xi)
\\&= \int_{v_i} z_i y_i \frac{p_{\mathcal E}(z_i)}
{p_{\mathcal O}(z_i \mid x_i, v; \xi)} p_{\mathcal O}(v_i \mid v; \xi) \d v_i \\
&= \int_{y_i, z_i, x_i} z_i y_i \frac{p_{\mathcal E}(z_i)}
{p_{\mathcal O}(z_i \mid x_i, v; \xi)} p_{\mathcal O}(y_i \mid z_i, x_i, v; \xi)
p_{\mathcal O}(z_i, x_i \mid v; \xi) \d y_i \d z_i \d x_i\\
&= \int_{z_i, x_i} z_i m(z_i, x_i; \xi) \frac{p_{\mathcal E}(z_i)}
{p_{\mathcal O}(z_i \mid x_i, v; \xi)} \sum_{k=1}^n \xi_k \delta_{(z_k, x_k)}(z_i, x_i) \d z_i \d x_i \\
&= \sum_{k=1}^n \xi_k z_k m(z_k, x_k; \xi) \frac{p_{\mathcal E}(z_k)}
{p_{\mathcal O}(z_i \mid x_k, v; \xi)} \\
&= \P_{\mathcal E}(Z_k = 1) \sum_{k=1}^n \xi_k  \frac{z_k m(z_k, x_k; \xi)}
{\P_{\mathcal O}(Z_k = 1 \mid x_k, v; \xi)},
\end{align*}
and therefore
\begin{align*}
\MoveEqLeft \E_{\mathcal E}\left\{\left.\frac{Z_i Y_i}{\P_{\mathcal E}(Z_i = 1)} - \frac{(1 - Z_i) 
Y_i}{\P_{\mathcal E}(Z_i = 0)} \,\right|\, v; \xi\right\} \\
&= \sum_{k=1}^n \xi_k m(z_k, x_k; \xi) \left\{\frac{z_k} {\P_{\mathcal O}(Z_k = 1 \mid x_k, v; \xi)} - \frac{1 - z_k}{\P_{\mathcal 
O}(Z_k = 0 \mid x_k, v; \xi)}\right\}.
\end{align*}
Thus, the posterior predictive mean difference can be written as
\begin{align}\label{equation:dr}
\MoveEqLeft \E_{\mathcal E}(Y_i \mid Z_i = 1, v; \xi) - \E_{\mathcal E}(Y_i \mid Z_i = 0, v; \xi) \nonumber \\
&= \E_{\mathcal E}\left\{\left.\frac{Z_i Y_i}{\P_{\mathcal E}(Z_i = 1)} - \frac{(1 - Z_i) Y_i}{\P_{\mathcal E}(Z_i 
= 0)} \,\right|\, v; \xi\right\}+ \E_{\mathcal E}(Y_i \mid Z_i = 1, v; \xi) \nonumber \\
&\quad
 - \E_{\mathcal E}\left\{\left.\frac{Z_i Y_i}{\P_{\mathcal E}(Z_i = 1)} - \frac{(1 - Z_i) Y_i}{\P_{\mathcal E}(Z_i 
= 0)} \,\right|\, v; \xi\right\} - \E_{\mathcal E}(Y_i \mid Z_i = 0, v; \xi) \nonumber \\
&= \sum_{k=1}^n \xi_k \{y_i - m(z_k, x_k; \xi)\} \left\{\frac{z_k} {\P_{\mathcal O}(Z_k = 1 \mid x_k, v; \xi)} - \frac{1 - 
z_k}{\P_{\mathcal O}(Z_k = 0 \mid x_k, v; \xi)}\right\} \nonumber \\
&\quad+ \sum_{k=1}^n \xi_k \left\{m(1, x_k; \xi) - m(0, x_k; \xi)\right\}.
\end{align}

\bibliographystyle{apalike}
{\small 
\setlength{\bibsep}{0pt plus 0.3ex}
\bibliography{ps}
}
\end{document}